\documentclass[epj]{svjour}
\usepackage{graphics}
\begin{document}
\title{Structure of the nucleon in chiral perturbation theory}
\author{Thomas Fuchs\inst{1}
\thanks{Supported by the Deutsche Forschungsgemeinschaft (SFB 443)}
\and Jambul Gegelia\inst{1}\inst{2}
\thanks{Alexander von Humboldt Research Fellow}
\and Stefan Scherer\inst{1}
}                     
\institute{Institut f\"ur Kernphysik, Johannes 
Gutenberg-Universit\"at, D-55099 Mainz, Germany 
\and 
High Energy Physics Institute, 
Tbilisi State University, 
University St.~9, 380086 Tbilisi, Georgia}
\date{Received: date / Revised version: date}
%
\abstract{
   We discuss a renormalization scheme for relativistic baryon
chiral perturbation theory which provides a simple and consistent power 
counting for renormalized diagrams.
   The method involves finite subtractions of dimensionally regularized 
diagrams beyond the standard modified minimal subtraction scheme of chiral 
perturbation theory to remove contributions violating the power counting.
   This is achieved by a suitable renormalization of the parameters
of the most general effective Lagrangian.
   As applications we discuss the mass of the nucleon, the $\sigma$ term,
and the scalar and electromagnetic form factors.
} 
\maketitle
\section{Introduction}
\label{introduction}
   Starting from Weinberg's pioneering work \cite{Weinberg:1979kz}, 
the application of effective field theory (EFT) to strong interaction 
processes has become one of the most important theoretical tools
in the low-energy regime.
   The basic idea consists of writing down the most general possible 
Lagrangian, including {\em all} terms consistent with assumed symmetry 
principles, and then calculating matrix elements with this Lagrangian 
within some perturbative scheme \cite{Weinberg:1979kz}.
   A successful application of this program thus requires two main ingredients:
\begin{enumerate}
\item[(1)]
a knowledge of the most general effective Lagrangian;
\item[(2)]
an expansion scheme for observables in terms of a consistent power counting
method.
\end{enumerate}
   The structure of the most general Lagrangian for both mesonic
and baryonic chiral perturbation theory (ChPT) has been investigated for
almost two decades.
   The number of terms in the momentum and quark-mass
expansion is given by
\begin{displaymath}
\underbrace{2}_{\mbox{${\cal O}(q^2)$}}
+\underbrace{10+2}_{\mbox{${\cal O}(q^4)$}}
+\underbrace{90+4+23}_{\mbox{${\cal O}(q^6)$}}
+\cdots
\end{displaymath}
for mesonic ChPT [SU(3)$\times$SU(3)] 
\cite{GL,p6} and
\begin{displaymath}
\underbrace{2}_{\mbox{${\cal O}(q)$}}
+\underbrace{7}_{\mbox{${\cal O}(q^2)$}}
+\underbrace{23}_{\mbox{${\cal O}(q^3)$}}
+\underbrace{118}_{\mbox{${\cal O}(q^4)$}}
+\cdots
\end{displaymath}
for baryonic ChPT [SU(2)$\times$SU(2)$\times$U(1)]
\cite{Gasser:1988rb,Krause:xc,Ecker:1995rk,Meissner:1998rw,Fettes:2000gb}.
   Moreover, the mesonic sector contains at ${\cal O}(q^4)$ the
Wess-Zumino-Witten action  \cite{Wess:yu,Witten:tw}
taking care of chiral anomalies.

  Once the most general effective Lagrangian is known, one needs an
expansion scheme in order to perform perturbative calculations of
physical observables.
   In this context one faces the standard difficulties of encountering
ultraviolet divergences when calculating loop diagrams.
   However, since one is working with the most general Lagrangian containing
all terms allowed by the symmetries, these infinities can, as part of
the renormalization program, be absorbed by a suitable adjustment of the 
parameters of the Lagrangian \cite{Weinberg:1979kz,Weinberg:mt}.
   Applying dimensional regularization in combination with the modified
minimal subtraction scheme of ChPT, in the mesonic sector a straightforward 
correspondence between the loop expansion and the chiral expansion 
in terms of momenta and quark masses at a fixed ratio was set up by 
Gasser and Leutwyler \cite{GL}. 
   The situation in the one-nucleon sector turned out to be more complicated
\cite{Gasser:1988rb}, since the correspondence between the loop expansion and 
the chiral expansion seemed to be lost.
   One of the findings of Ref.\ \cite{Gasser:1988rb} was that higher-loop 
diagrams can contribute to terms as low as ${\cal O}(q^2)$.
   A solution to this problem was obtained in the framework of the 
heavy-baryon formulation of ChPT \cite{Jenkins:1990jv,Bernard:1992qa} 
resulting in a power counting analogous to the mesonic sector
(for a recent review of ChPT see, e.g., Ref.\ \cite{Scherer:2002tk}).

   Here, we will review some recent efforts to devise a new renormalization 
scheme leading to a simple and consistent power counting for the renormalized 
diagrams of a manifestly covariant approach.
   The basic idea consists in performing additional subtractions of 
dimensionally regularized diagrams beyond the modified minimal subtraction 
scheme employed in Ref.\ \cite{Gasser:1988rb}.
   As applications we will discuss the mass of the nucleon as well
as the scalar and electromagnetic form factors and compare the method
with the approach of Becher and Leutwyler \cite{Becher:1999he}.

\section{Dimensional regularization}
\label{dimensional_regularization}
\begin{figure}
\begin{center}
\resizebox{0.1\textwidth}{!}{%
\includegraphics{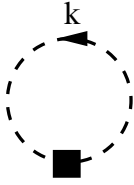}
}
\end{center}
\caption{Generic one-loop diagram. The black box denotes some unspecified
vertex structure which is irrelevant for the discussion.}
\label{fig:dimregexample}       
\end{figure}
   For the regularization of loop diagrams we will make use of dimensional
regularization \cite{dimreg},
because it preserves algebraic relations between Green
functions (Ward identities).
   We will illustrate the method by considering the following simple example,
\begin{equation}
\label{idef}
I(M^2)=
\int\frac{d^4k}{(2\pi)^4}\frac{i}{k^2-M^2+i0^+},\quad k^2=k_0^2-\vec{k}\,^2,
\end{equation}
which shows up in the generic diagram of Fig.\ \ref{fig:dimregexample}. 
   Naively counting the powers of the momenta, the integral is said
to diverge quadratically.
   In order to regularize Eq.\ (\ref{idef}), we define the integral for $n$ 
dimensions ($n$ integer) as 
\begin{displaymath}
I_n(M^2,\mu^2)=\mu^{4-n}\int\frac{d^nk}{(2\pi)^n}\frac{i}{k^2-M^2+i0^+},
\end{displaymath}
where the scale $\mu$ ('t Hooft parameter) has been introduced 
so that the integral has the same dimension for arbitrary $n$.
   After a Wick rotation and angular integration, the analytic continuation 
for complex $n$ reads (see Appendix B of Ref.\ \cite{Scherer:2002tk} for
details)
\begin{eqnarray}
\label{I}
I(M^2,\mu^2,n)&=&
\frac{M^2}{(4\pi)^2}\left(\frac{4\pi\mu^2}{M^2}\right)^{2-\frac{n}{2}}
\Gamma\left(1-\frac{n}{2}\right)\nonumber\\
&=&\frac{M^2}{16\pi^2}\left[
R+\ln\left(\frac{M^2}{\mu^2}\right)\right]+O(n-4),
\end{eqnarray}
where
\begin{equation}
\label{R}
R=\frac{2}{n-4}
-[\mbox{ln}(4\pi)+\Gamma'(1)]-1.
\end{equation}
   The idea of renormalization consists of adjusting the parameters of the 
counterterms of the most general effective Lagrangian 
so that they cancel the divergences of (multi-) loop diagrams.
   In doing so, one still has the freedom of choosing a suitable 
renormalization condition.
   For example, in the minimal subtraction scheme (MS) one would fix the 
parameters of the counterterm Lagrangian such that they would precisely absorb 
the contributions proportional to $2/(n-4)$ in Eq.\ (\ref{R}), 
while the modified minimal subtraction scheme ($\overline{\rm MS}$) would, 
in addition, cancel the term in square brackets.
   Finally, in the modified minimal subtraction scheme of ChPT
($\widetilde{\rm MS}$) employed in Ref.\ \cite{GL}, 
the seven (bare) coefficients $l_i$ of the ${\cal O}(q^4)$ Lagrangian are 
expressed in terms of renormalized coefficients $l_i^r$ as
\begin{equation}
\label{li}
l_i=l_i^r+\gamma_i\frac{R}{32\pi^2},
\end{equation}
where the $\gamma_i$ are fixed numbers.

\section{Mesonic chiral perturbation theory}
\label{mesonic_chpt}
   The starting point of mesonic chiral perturbation theory is a chiral
$\mbox{SU(2)}_L\times\mbox{SU(2)}_R$ symmetry of the two-flavor QCD 
Lagrangian in the limit of massless $u$ and $d$ quarks. 
   It is assumed that this symmetry is spontaneously broken down
to its isospin subgroup SU(2)$_V$, i.e., the ground state has a
lower symmetry than the Lagrangian. 
   From Goldstone's theorem one expects $6-3 =3$ massless Goldstone 
bosons which interact ``weakly'' at low energies, and which are identified
with the pions of the ``real'' world.
   The explicit chiral symmetry breaking through the quark masses is
included as a perturbation.
   According to the program of EFT the symmetries of
QCD are mapped onto the most general effective Lagrangian for the 
interaction of the Goldstone bosons (pions).
   The Lagrangian is organized in a derivative
and quark-mass expansion
\cite{Weinberg:1979kz,GL,p6}
\begin{equation}
\label{lpi}
{\cal L}_\pi={\cal L}_2 + {\cal L}_4 + {\cal L}_6 + \cdots,
\end{equation}
where---in the absence of external fields---the lowest-order Lagrangian is 
given by \cite{GL} 
\begin{equation}
\label{l2}
{\cal L}_2 = \frac{F^2}{4} \mbox{Tr} \left( \partial_{\mu} U 
\partial^{\mu}U^{\dagger}\right)
+\frac{F^2 M^2}{4}\mbox{Tr}(U^{\dagger}+ U),
\end{equation}
with 
\begin{displaymath}
U=\exp\left(i\frac{\vec{\tau}\cdot\vec{\pi}}{F}\right)
\end{displaymath}
a unimodular unitary $(2\times 2)$ matrix containing 
the Goldstone boson fields. 
   In Eq.\ (\ref{l2}), $F$ denotes the pion-decay constant in the chiral 
limit: $F_\pi=F[1+{\cal O}(\hat{m})]=92.4$ MeV. 
   Here, we work in the isospin-symmetric limit $m_u=m_d=\hat{m}$, 
and the lowest-order expression for the squared pion mass is 
$M^2=2 B \hat{m}$, where $B$ is related to the quark condensate 
$\langle \bar{q} q\rangle_0$ in the chiral limit \cite{GL}.
   
   Using Weinberg's power counting scheme \cite{Weinberg:1979kz}
one may analyze the behavior
of a given diagram calculated in the framework of Eq.\ (\ref{lpi}) under
a linear rescaling of all {\em external} momenta, $p_i\mapsto t p_i$, 
and a quadratic rescaling of the light quark masses, $m_q\mapsto t^2 m_q$, 
which, in terms of the Goldstone boson masses, corresponds to 
$M^2\mapsto t^2 M^2$.
   The chiral dimension $D$ of a given diagram with amplitude
${\cal M}(p_i,m_q)$ is defined by
\begin{equation}
\label{mr1}
{\cal M}(tp_i, t^2 m_q)=t^D {\cal M}(p_i,m_q),
\end{equation}
where, in $n$ dimensions, 
\begin{eqnarray}
D&=&n N_L-2I_\pi+\sum_{k=1}^\infty 2k N_{2k}^\pi\label{mr2a}\\
&=&2+(n-2) N_L+\sum_{k=1}^\infty 2(k-1) N_{2k}^\pi \label{mr2b}\\
&\geq&\mbox{2 in 4 dimensions}.\nonumber
\end{eqnarray}
   Here, $N_L$ is the number of independent loop momenta, 
$I_\pi$ the number of internal pion lines,
and $N_{2k}^\pi$ the number of vertices originating from ${\cal L}_{2k}$.
   Clearly, for small enough momenta and masses diagrams with small $D$, such 
as $D=2$ or $D=4$, should dominate.
   Of course, the rescaling of Eq.\ (\ref{mr1}) must be viewed as 
a mathematical tool.
   While external three-momenta can, to a certain extent, be made arbitrarily
small, the rescaling of the quark masses is a theoretical instrument only.
   Note that, for $n=4$, loop diagrams are always suppressed due to the term 
$2N_L$ in Eq.\ (\ref{mr2b}).    
   In other words, we have a perturbative scheme in terms of external 
momenta and masses which are small compared to some scale
[here $1/(4\pi F)$].

\begin{figure}
\begin{center}
\resizebox{0.2\textwidth}{!}{%
\includegraphics{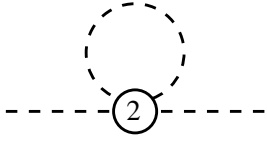}
}
\end{center}
\caption{One-loop contribution to the pion self-energy. The number 2 in the 
interaction blob refers to ${\cal L}_2$.}
\label{fig:examplepionselfenergy}       
\end{figure}

   As an example, let us consider the contribution of 
Fig.~\ref{fig:examplepionselfenergy} to the pion self-energy.
   According to Eq.\ (\ref{mr2a}) we expect, in 4 dimensions, the chiral power
\begin{displaymath}
D=4\cdot 1-2\cdot 1+ 2 \cdot 1=4.
\end{displaymath}
   Without going into the details, the explicit result of the 
one-loop contribution is given by (see, e.g., Ref.\ \cite{Scherer:2002tk})
\begin{displaymath}
\Sigma_{\rm loop}(p^2)=
\frac{4p^2-M^2}{6 F^2}
I(M^2,\mu^2,n)
={\cal O}(q^4),
\end{displaymath}
   where the integral is given in Eq.\ (\ref{I}) and is infinite as 
$n\to 4$.
   Note that both factors---the fraction and the integral---each count as
${\cal O}(q^2)$ resulting in ${\cal O}(q^4)$ for the total expression
as anticipated.

   For a long time it was believed that performing loop 
calculations using the Lagrangian of Eq.\ (\ref{l2}) would make no sense,
because it is not renormalizable (in the traditional sense).
   However, as emphasized by Weinberg  \cite{Weinberg:1979kz,Weinberg:mt},  
the cancellation of ultraviolet divergences does not really depend on
renormalizability; as long as one includes \mbox{\em every one} of the infinite
number of interactions allowed by symmetries, the so-called non-renormalizable
theories are actually just as renormalizable as renormalizable theories
\cite{Weinberg:mt}.
   The conclusion is that a suitable adjustment of the parameters of  
${\cal L}_4$ [see Eq.\ (\ref{li})] leads to a cancellation of the one-loop 
infinities.

\section{Baryonic chiral perturbation theory and renormalization} 
\label{baryonic_chpt}
   The extension to processes involving one external nucleon line was 
developed by Gasser, Sainio, and \v{S}varc \cite{Gasser:1988rb}.
   In addition to Eq.\ (\ref{lpi}) one needs the most general effective
Lagrangian of the interaction of Goldstone bosons with nucleons:
\begin{displaymath}
{\cal L}_{\pi N}={\cal L}_{\pi N}^{(1)}+{\cal L}_{\pi N}^{(2)}+\cdots.
\end{displaymath}
   The lowest-order Lagrangian,
expressed in terms of bare fields and parameters denoted by subscripts 0,
reads 
\begin{equation} 
\label{lpin1}
{\cal L}_{\pi N}^{(1)}=\bar \Psi_0 \left( i\gamma_\mu 
\partial^\mu -m_0 -\frac{1}{2}\frac{{\stackrel{\circ}{g_{A}}}_0}{F_0} 
\gamma_\mu 
\gamma_5 \tau^a \partial^\mu \pi^a_0\right) \Psi_0 +\cdots,
\end{equation} 
   where $\Psi_0$ denotes the (bare) nucleon field with two four-component 
Dirac fields describing the proton and the neutron, respectively.
   After renormalization, $m$ and $\stackrel{\circ}{g_A}$ refer to the chiral 
limit of the physical nucleon mass and the axial-vector coupling constant, 
respectively. 
   While the mesonic Lagrangian of Eq.\ (\ref{lpi}) contains only even powers, 
the baryonic Lagrangian involves both even and odd powers due to the 
additional spin degree of freedom.  

   Our goal is to propose a renormalization procedure generating a power 
counting for tree-level and loop diagrams of the (relativistic) EFT 
which is analogous to that given in Ref.\ \cite{Weinberg:1991um} 
(for nonrelativistic nucleons). 
   Choosing a suitable renormalization condition will
allow us to apply the following power counting: 
a loop integration in $n$ dimensions counts as $q^n$, 
pion and fermion propagators count as $q^{-2}$ and 
$q^{-1}$, respectively, vertices derived from ${\cal L}_{2k}$ and 
${\cal L}_{\pi N}^{(k)}$ count as $q^{2k}$ and $q^k$, respectively.
   Here, $q$ generically denotes a small expansion parameter such as,
e.g., the pion mass.
   In total this yields for the power $D$ of a diagram in the 
one-nucleon sector the standard formula 
\cite{Weinberg:1991um,Ecker:1995gg}
\begin{eqnarray}
\label{dimension1}
D&=&n N_L - 2 I_\pi - I_N +\sum_{k=1}^\infty 2k N^\pi_{2k}
+\sum_{k=1}^\infty k N_k^N\\
\label{dimension2}
&=&1+(n-2)N_L+\sum_{k=1}^\infty 2(k-1) N^\pi_{2k}
+\sum_{k=1}^\infty (k-1) N_k^N\nonumber\\ &&\\
&\geq&\mbox{1 in 4 dimensions},\nonumber
\end{eqnarray}
where, in addition to Eq.\ (\ref{mr2a}), $I_N$ is the number of internal 
nucleon lines and $N_k^N$ the number of vertices originating from 
${\cal L}_{\pi N}^{(k)}$.
   According to Eq.\ (\ref{dimension2}), one-loop calculations in the
single-nucleon sector should
start contributing at ${\cal O}(q^{n-1})$.

\begin{figure}
\begin{center}
\resizebox{0.3\textwidth}{!}{%
\includegraphics{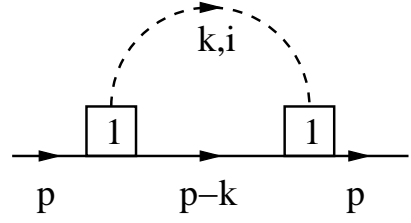}
}
\end{center}
\caption{One-loop contribution to the nucleon self-energy.
The number 1 in the interaction blobs refers to ${\cal L}_{\pi N}^{(1)}$.}
\label{fig:nucleonselfenergypionloop}       
\end{figure}

   As an example, let us consider the one-loop contribution of 
Fig.\ \ref{fig:nucleonselfenergypionloop} to the nucleon self-energy.
   According to Eq.\ (\ref{dimension1}), the renormalized result should 
be of order
\begin{equation}
\label{dexample}
D=n\cdot 1-2\cdot 1-1\cdot 1+1\cdot 2=n-1.
\end{equation}
   An explicit calculation yields 
\begin{eqnarray*} 
\label{sigmaaresult} 
\lefteqn{\Sigma_{\rm loop}=
-\frac{3 {\stackrel{\circ}{g_{A}}}_0^2}{4 F_0^2}\left\{ 
\vphantom{-\frac{(p^2-m^2)p\hspace{-.5em}/}{2p^2}} 
(p\hspace{-.45em}/\hspace{.1em}+m)I_N 
+M^2(p\hspace{-.45em}/\hspace{.1em}+m)I_{N\pi}(-p,0)\right.}\nonumber\\ 
&&\left. -\frac{(p^2-m^2)p\hspace{-.45em}/\hspace{.1em}}{2p^2}[ 
(p^2-m^2+M^2)I_{N\pi}(-p,0)+I_N-I_\pi]\right\}, 
\end{eqnarray*} 
where the relevant loop integrals are defined as
\begin{eqnarray}
\label{Ipi}
I_\pi&=& \mu^{4-n}\int\frac{d^nk}{(2\pi)^n}\frac{i}{k^2-M^2+i0^+},\\
\label{IN}
I_N&=& \mu^{4-n}\int\frac{d^nk}{(2\pi)^n}\frac{i}{k^2-m^2+i0^+},\\
\label{INpi}
I_{N\pi}(-p,0)&=&\mu^{4-n}\int\frac{d^nk}{(2\pi)^n}
\frac{i}{[(k-p)^2-m^2+i0^+]}\nonumber\\
&&\hspace{2em}\times\frac{1}{k^2-M^2+i0^+}.
\end{eqnarray}
   Applying the $\widetilde{\rm MS}$ renormalization scheme---indicated 
by ``r''---one obtains 
\begin{displaymath} 
\Sigma_{\rm loop}^r=-\frac{3 g_{Ar}^2}{4 F_r^2}\left[ 
-\frac{M^2}{16\pi^2}(p\hspace{-.4em}/\hspace{.1em}+m)
+\cdots\right]=
{\cal O}(q^2),
\end{displaymath}
   i.e., the $\widetilde{\rm MS}$-renormalized result does not produce
the desired low-energy behavior of Eq.\ (\ref{dexample}).
   Gasser, Sainio, and \v{S}varc concluded that loops have a much more 
complicated low-energy structure if baryons are included. 
   The appearance of another scale, namely, the mass of the nucleon 
(which does not vanish in the chiral limit), is one of the origins for the 
complications in the baryonic sector \cite{Gasser:1988rb}. 
   The apparent ``mismatch'' between the chiral and the loop expansion has 
widely been interpreted as the absence of a systematic power counting
in the relativistic formulation.

\subsection{Heavy-baryon approach}
\label{hb}
   One possibility of overcoming the problem of power counting was provided 
by the heavy-baryon formulation of ChPT \cite{Jenkins:1990jv,Bernard:1992qa} 
resulting in a power counting scheme which follows Eqs.\ (\ref{dimension1})
and (\ref{dimension2}).
   The basic idea consists in dividing nucleon momenta into a large piece 
close to on-shell kinematics and a soft residual contribution: 
$p = m v +k_p$, $v^2=1$, $v^0\ge 1$
[often $v^\mu = (1,0,0,0)$].
   The relativistic nucleon field is expressed in terms of 
velocity-dependent fields,
\begin{displaymath}
\Psi(x)=e^{-im v \cdot x} ({\cal N}_v +{\cal H}_v), 
\end{displaymath}
with 
\begin{displaymath}
{\cal N}_v=e^{+im v\cdot x}\frac{1}{2}(1+v\hspace{-.5em}/)\Psi,\quad
{\cal H}_v=e^{+im v\cdot x}\frac{1}{2}(1-v\hspace{-.5em}/)\Psi.
\end{displaymath}
   Using the equation of motion for ${\cal H}_v$, one can
eliminate ${\cal H}_v$ and obtain a Lagrangian for ${\cal N}_v$
which, to lowest order, reads \cite{Bernard:1992qa}
\begin{displaymath}
\widehat{\cal L}^{(1)}_{\pi N}=\bar{\cal N}_v(iv\cdot D + g_A S_v\cdot u)
{\cal N}_v+{\cal O}(1/m).
\end{displaymath}
   The result of the heavy-baryon reduction is a $1/m$ expansion of the 
Lagrangian similar to a Foldy-Wouthuysen expansion.  
   Now, power counting works along Eqs.\ (\ref{dimension1})
and (\ref{dimension2}) but the approach has its own shortcomings.
   In higher orders in the chiral expansion, the expressions due to
$1/m$ corrections of the Lagrangian become increasingly complicated.

\begin{figure}
\begin{center}
\resizebox{0.3\textwidth}{!}{%
\includegraphics{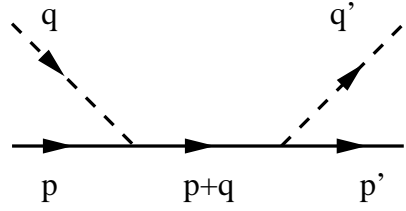}
}
\end{center}
\caption{$s$-channel pole diagram of $\pi$N scattering.}
\label{fig:schannel}       
\end{figure}
   Moreover---and what is more important---the approach generates problems
regarding analyticity.
   This can easily be illustrated by considering the example of
pion-nucleon scattering \cite{Becher:2000mb}.
   The invariant amplitudes describing the scattering amplitude develop
poles for $s=m_N^2$ and $u=m_N^2$.
   For example, the singularity due to the nucleon pole in the $s$ channel
(see Fig.\ \ref{fig:schannel}) is understood in terms of the relativistic 
propagator
\begin{equation}
\label{relprop}
\frac{1}{(p+q)^2-m_N^2}=\frac{1}{2p\cdot q+M_\pi^2},
\end{equation}
which, of course, has a pole at $2p\cdot q=-M_\pi^2$ or, equivalently,
$s=m_N^2$.
   (Analogously, a second pole results from the $u$ channel at $u=m_N^2$.)
   Although both poles are not in the physical region of pion-nucleon 
scattering, analyticity of the invariant amplitudes requires these poles
to be present in the amplitudes.
   Let us compare the situation with a heavy-baryon type of expansion,
where, for simplicity, we choose as the four-velocity $p^\mu=m_N v^\mu$,
\begin{eqnarray}
\label{relpropv}
\frac{1}{2p\cdot q+M_\pi^2}&=&
\frac{1}{2m_N}\frac{1}{v\cdot q+ \frac{M_\pi^2}{2m_N}}\nonumber\\
&=&\frac{1}{2 m_N}\frac{1}{v\cdot q}\left(1-\frac{M_\pi^2}{2 m_N v\cdot q}
+\cdots\right).
\end{eqnarray}
   Clearly, to any finite order the heavy-baryon expansion produces poles 
at $v\cdot q=0$ instead of a simple pole at $v\cdot q=-M_\pi^2/(2 m_N)$ and 
will thus not generate the (nucleon) pole structures of the invariant 
amplitudes. 
   Another example involving loop diagrams will be given in 
Section \ref{applications}.
   
\subsection{Infrared regularization}

   A second solution was offered by Becher and Leutwyler \cite{Becher:1999he} 
and is referred to as the so-called infrared regularization.
   The basic idea can be illustrated using the loop integral 
of Eq.\ (\ref{INpi}).
   To that end, we make use of the Feynman parametrization 
\begin{displaymath} 
{1\over ab}=\int_0^1 {dz\over [az+b(1-z)]^2} 
\end{displaymath} 
with $a=(k-p)^2-m^2+i0^+$ and $b=k^2-M^2+i0^+$, interchange the 
order of integrations, and perform the shift $k\to k+zp$.
   The resulting integral over the Feynman parameter $z$ is then rewritten as 
\begin{eqnarray*}
I_{N\pi}(-p,0)=\int_0^1 dz \cdots &=& \int_0^\infty dz \cdots 
- \int_1^\infty dz \cdots,\\
\end{eqnarray*}
where the first, so-called infrared (singular) integral satisfies the power
counting, while the remainder violates power counting but turns out to
be regular and can thus be absorbed in counterterms.
   In the one-nucleon sector, it is straightforward to generalize the method 
for any one-loop integral consisting of an arbitrary number of nucleon and 
pion propagators \cite{Becher:1999he} (see also Ref.\ \cite{Goity:2001ny}).

\subsection{Extended on-mass-shell scheme}
   In the following, we will concentrate on yet another solution which 
has been motivated in Ref.\ \cite{Gegelia:1999gf} and has been worked out in 
detail in Ref.\ \cite{Fuchs:2003qc} (for other approaches, see Refs.\ 
\cite{otherapproaches}).
   The central idea consists of performing additional subtractions beyond
the $\widetilde{\rm MS}$ scheme such that 
renormalized diagrams satisfy the power counting.
   Terms violating the power counting are analytic in small
quantities and can thus be absorbed in a renormalization of 
counter\-terms.
   In order to illustrate the approach, let us consider as an example
the integral 
\begin{displaymath}
H(p^2,m^2;n)= \int \frac{d^n k}{(2\pi)^n}
\frac{i}{[(k-p)^2-m^2+i0^+][k^2+i0^+]},
\end{displaymath}
where 
\begin{displaymath}
\Delta=\frac{p^2-m^2}{m^2}={\cal O}(q)
\end{displaymath}
is a small quantity.
   We want the (renormalized) integral to be of order
\begin{displaymath}
D=n-1-2=n-3.
\end{displaymath}
   The result of the integration is of the form (see Ref.\ 
\cite{Fuchs:2003qc} for details)
\begin{displaymath}
H\sim F(n,\Delta)+\Delta^{n-3}G(n,\Delta),
\end{displaymath}
where $F$ and $G$ are hypergeometric functions and are analytic in $\Delta$ for
any $n$.
   Hence, the part containing $G$ for noninteger $n$ is proportional to
a noninteger power of $\Delta$ and satisfies the power counting.
   The part proportional to $F$ can be obtained by first expanding
the integrand in small quantities and {\em then} performing the integration
 for each term \cite{Gegelia:1994zz}.
   It is this part which violates the power counting, but, since
it is analytic in $\Delta$, the power-counting violating pieces can
be absorbed in the counterterms.
   This observation suggests the following procedure: expand the integrand in 
small quantities and subtract those (integrated) terms whose order is 
smaller than suggested by the power counting.
   In the present case, the subtraction term reads
\begin{displaymath}
H^{\rm subtr}=\int \frac{d^n k}{(2\pi)^n}\left.
\frac{i}{[k^2-2p\cdot k +i0^+][k^2+i0^+]}\right|_{p^2=m^2}
\end{displaymath}
and the renormalized integral is written as
\begin{displaymath}
H^R=H-H^{\rm subtr}={\cal O}(q^{n-1}).
\end{displaymath}
   Using our EOMS scheme it is also possible to include (axial) vector mesons
explicitly \cite{Fuchs:2003sh}.
   Moreover, the infrared regularization of Becher and Leutwyler
can be reformulated in a form analogous to the EOMS renormalization
scheme and can thus be applied straightforwardly to multi-loop diagrams
with an arbitrary number of particles with arbitrary masses 
\cite{Schindler:2003xv} (see also Ref.\ \cite{Goity:2001ny}).

\section{Applications}
\label{applications}
   As the first application, we discuss the result for the mass of the nucleon
at ${\cal O}(q^3)$. 
   Within the $\widetilde{\rm MS}$ scheme of Ref.\ \cite{Gasser:1988rb}
the result is given by
\begin{equation}
\label{mmstilde}
m_N=m-4 c_1^r M^2+\frac{3 g_{Ar}^2 M^2}{32 \pi^2 F_r^2}m 
\left(1+8 c_1^r m\right)
-\frac{3 g_{Ar}^2 M^3}{32 \pi F_r^2},
\end{equation}
where $r$ indicates $\widetilde{\rm MS}$-renormalized quantities, and where
we have used the renormalization scale $\mu=m$ with $m$ the $\mbox{SU(2)}
\times\mbox{SU(2)}$ chiral limit of the nucleon mass (at fixed $m_s\neq 0$). 
   The third term on the r.~h.~s.~of Eq.\ (\ref{mmstilde}) violates the
power counting of Eq.\ (\ref{dimension1}), because it is proportional to
$M^2$, i.e., ${\cal O}(q^2)$, 
while it is obtained from the diagram of Fig.\ 
\ref{fig:nucleonselfenergypionloop} which should generate contributions
of ${\cal O}(q^3)$.
   On the other hand, the result in the EOMS scheme is given by 
\cite{Fuchs:2003qc}
\begin{equation}
m_N=m-4 c_1 M^2-\frac{3 g_A^2 M^3}{32 \pi F^2}+{\cal O}(M^4),
\end{equation}  
where all parameters are understood to be taken in the EOMS scheme.
   Clearly, this expression satisfies the power counting, because the
renormalized loop contribution of Fig.\ \ref{fig:nucleonselfenergypionloop}
is of ${\cal O}(M^3)$.
   The relation between the $\widetilde{\rm MS}$-renormal\-ized 
and the EOMS-renormalized coefficients is given by
\begin{displaymath} 
c_{1}^r=c_1+\frac{3 m g_A^2}{128 \pi^2 F^2}\left[ 1+8 mc_1 \right]+\cdots. 
\end{displaymath} 

   A full calculation of the nucleon mass at ${\cal O}(q^4)$ yields  
\cite{Fuchs:2003qc}
\begin{equation}
\label{mnoq4} 
m_N=m+k_1 M^2+k_2 M^3+k_3 M^4\ln\left(\frac{M}{m}\right)+k_4 M^4
+{\cal O}(M^5), 
\end{equation}
where the coefficients $k_i$ are given by
\begin{eqnarray} 
\label{parki} 
k_1&=&-4 c_1,\quad 
k_2=-\frac{3 {\stackrel{\circ}{g_A}}^2}{32\pi F^2},\nonumber\\ 
k_3&=&\frac{3}{32\pi^2 F^2}\left(8c_1-c_2-4 c_3
-\frac{{\stackrel{\circ}{g_A}}^2}{m}\right), 
\nonumber\\ 
k_4&=&\frac{3 {\stackrel{\circ}{g_A}}^2}{32 \pi^2 F^2 m}(1+4 c_1 m) 
+\frac{3}{128\pi^2F^2}c_2+\frac{1}{2}\alpha.
\end{eqnarray}
Here, $\alpha=-4(8 e_{38}+e_{115}+e_{116})$ is a linear combination
of ${\cal O}(q^4)$ coefficients \cite{Fettes:2000gb}.
  In order to obtain an estimate for the various contributions 
of Eq.\ (\ref{mnoq4}) to the nucleon mass, we make use of 
the set of parameters $c_i$ of Ref.\ \cite{Becher:2001hv}, 
\begin{eqnarray}
\label{parametersci} 
&&c_1=-0.9\,m_N^{-1},\quad 
c_2=2.5\, m_N^{-1},\nonumber\\ 
&&c_3=-4.2\, m_N^{-1},\quad 
c_4=2.3\, m_N^{-1}. 
\end{eqnarray} 
   These numbers were obtained from a (tree-level) fit to the $\pi N$ 
scattering threshold parameters of Ref.\ \cite{Koch:bn}. 
   Using the numerical values
\begin{eqnarray}
\label{numericalvalues} 
&&g_A=1.267,\quad F_\pi=92.4\,\mbox{MeV},\quad 
m_N=m_p=938.3\,\mbox{MeV},\nonumber\\
&&M_\pi=M_{\pi^+}=139.6\,\mbox{MeV}, 
\end{eqnarray} 
we obtain for the mass of nucleon in the chiral limit (at
fixed $m_s\neq 0$):
\begin{eqnarray*}
m&=&m_N-\Delta m\nonumber\\
&=&[938.3-74.8+15.3+4.7+1.6-2.3]\,\mbox{MeV}\\
&=&882.8\, \mbox{MeV}
\end{eqnarray*}
with $\Delta m=55.5\,\mbox{MeV}$. 
   Here, we have made use of an estimate for $\alpha$ obtained from
the $\sigma$ term (see the following discussion).

   Similarly, an analysis of the $\sigma$ term yields
\begin{equation} 
\sigma=\sigma_1 M^2+\sigma_2 M^3 +\sigma_3 M^4 \ln\left(\frac{M}{m}\right) 
+\sigma_4 M^4+{\cal O}(M^5), 
\end{equation}
with
\begin{eqnarray} 
\label{parsigmai} 
\sigma_1&=&-4 c_1,\quad 
\sigma_2=-\frac{9{\stackrel{\circ}{g_{A}}}^2}{64\pi F^2},\nonumber\\ 
\sigma_3&=&\frac{3}{16\pi^2 F^2}\left(8c_1-c_2-4c_3
-\frac{{\stackrel{\circ}{g_A}}^2}{m}\right), 
\nonumber\\ 
\sigma_4&=&\frac{3}{8\pi^2 F^2}\left[\frac{3 {\stackrel{\circ}{g_{A}}}^2}{8 m} 
+c_1(1+2 {\stackrel{\circ}{g_A}}^2)-\frac{c_3}{2}\right]+\alpha.
\end{eqnarray} 
   We obtain [with $\alpha=0$ in Eq.\ (\ref {parsigmai})] 
\begin{equation} 
\label{sigmanum} 
\sigma=(74.8-22.9-9.4-2.0)\,\mbox{MeV}=40.5\, \mbox{MeV}. 
\end{equation} 
   The result of Eq.\ (\ref{sigmanum}) has to be compared with the 
dispersive analysis $\sigma=(45\pm 8)$ MeV of Ref.\ \cite{Gasser:1990ce} 
which would imply, neglecting higher-order terms, $\alpha M^4 \approx 4.5$
MeV. 
   As has been discussed, e.g., in Ref.\ \cite{Becher:1999he},
a fully consistent description would also require to determine 
the low-energy coupling constant $c_1$ from a complete ${\cal O}(q^4)$ 
calculation of, say, $\pi N$ scattering. 

   The results of Eqs.\ (\ref{parki}) and (\ref{parsigmai}) satisfy the
constraints as implied by the application of the Hellmann-Feynman theorem 
to the nucleon mass \cite{GL,Gasser:1988rb} 
\begin{equation} 
\sigma =M^2 \ {\partial m_N\over \partial M^2}. 
\label{fhrelation} 
\end{equation} 

   A chiral low-energy theorem \cite{Cheng:mx,Brown:pn} relates the 
scalar form factor at $t=2M_\pi^2$ to the $\pi N$ scattering amplitude at the 
unphysical point $\nu=0$, $t=2M_\pi^2$ 
(for a recent discussion of the corrections, see Ref.\ \cite{Becher:2001hv}).
   Defining the difference $\Delta_\sigma=\sigma(2 M_\pi^2) -\sigma(0)$,
one obtains a similar expansion for $\Delta_\sigma$
as for the nucleon mass and the $\sigma$ term \cite{Becher:1999he} 
 \begin{equation} 
\label{deltasigmaparameterization} 
\Delta_\sigma=
\Delta_1 M^3+\Delta_2 M^4 \ln\left(\frac{M}{m}\right) 
+\Delta_3 M^4+{\cal O}(M^5), 
\end{equation}
where 
\begin{eqnarray} 
\label{pardeltasigmai} 
&&\Delta_1=\frac{3 {\stackrel{\circ}{g_{A}}}^2}{64\pi F^2},\quad
\Delta_2=\frac{1}{16\pi^2 F^2}\left(\frac{3{\stackrel{\circ}{g_{A}}}^2}{m}
+c_2+6c_3\right), 
\nonumber\\
&&\Delta_3=8 e_{22} -\frac{c_1 {\stackrel{\circ}{g_{A}}}^2}{4\pi^2 F^2}
+\frac{3(\pi-2){\stackrel{\circ}{g_{A}}}^2}{128\pi^2 F^2 m}
+\frac{3c_1(\pi-4)}{16\pi^2 F^2}\nonumber\\
&&\quad \quad\quad +\frac{c_2(14-3\pi)}{192\pi^2 F^2}
+\frac{3 c_3}{16\pi^2 F^2},
\end{eqnarray} 
where $e_{22}$ is an ${\cal O}(q^4)$ coefficient \cite{Fettes:2000gb}.
   Using the parameters and numerical values of Eqs.\ (\ref{parametersci})
and (\ref{numericalvalues}), respectively, we obtain [with $e_{22}=0$ 
in Eq.\ (\ref {pardeltasigmai})] 
\begin{equation} 
\label{deltasigmanum} 
\Delta_\sigma=(7.6+10.2-0.9)\,\mbox{MeV}=16.9\, \mbox{MeV}, 
\end{equation} 
which has to be compared with the dispersive analysis 
$\Delta_\sigma=(15.2\pm 0.4)$ MeV
of Ref.\ \cite{Gasser:1990ce} resulting in the estimate
$8 e_{22} M^4 \approx -1.7$ MeV.

\begin{figure}
\begin{center}
\resizebox{0.5\textwidth}{!}{%
\includegraphics{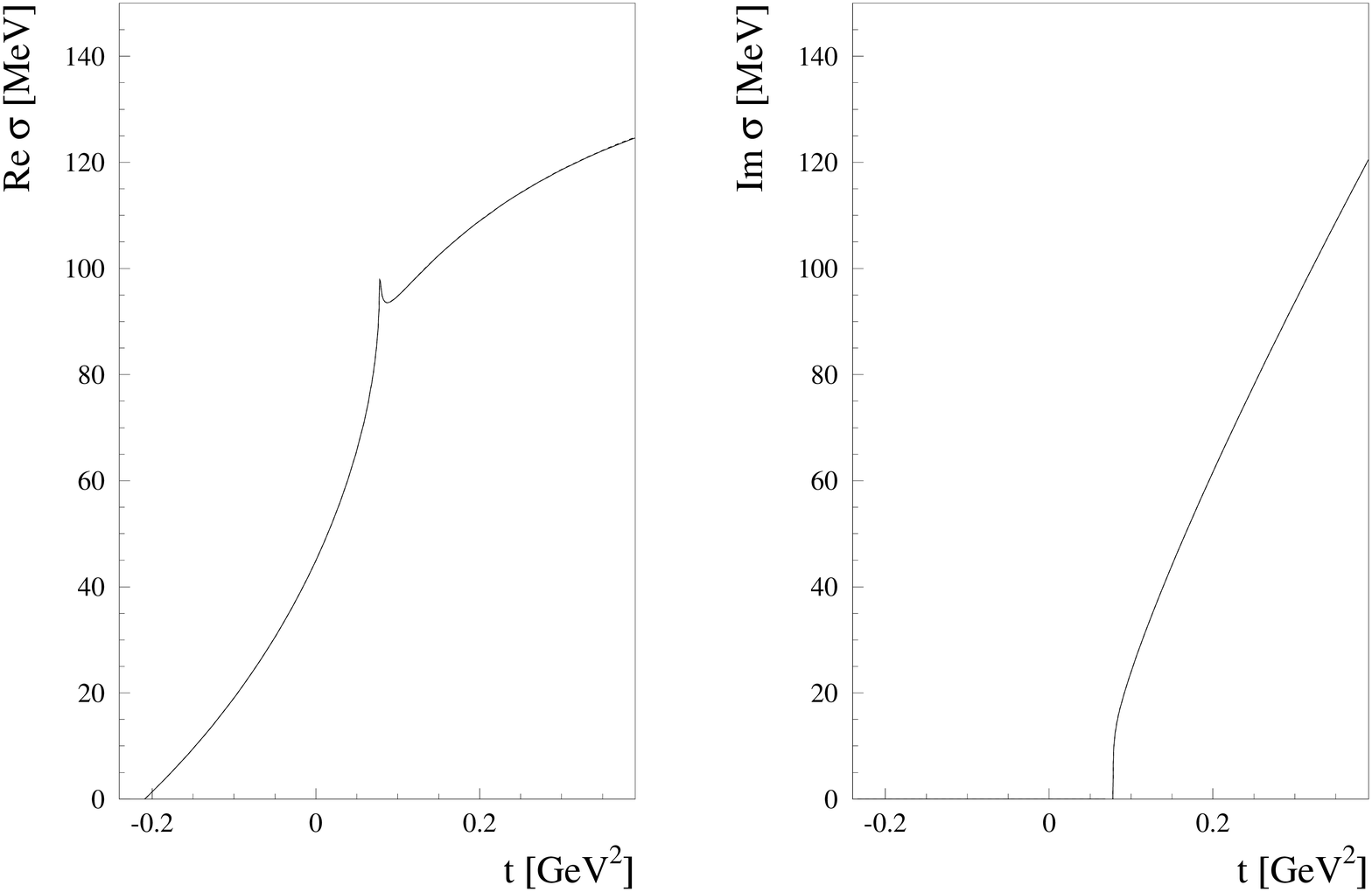}
}
\end{center}
\caption{Scalar form factor $\sigma(t)$ as a function of $t$ at 
${\cal O}(q^4)$.}
\label{fig:scffq4new}       
\end{figure}

\begin{figure}
\begin{center}
\resizebox{0.5\textwidth}{!}{%
\includegraphics{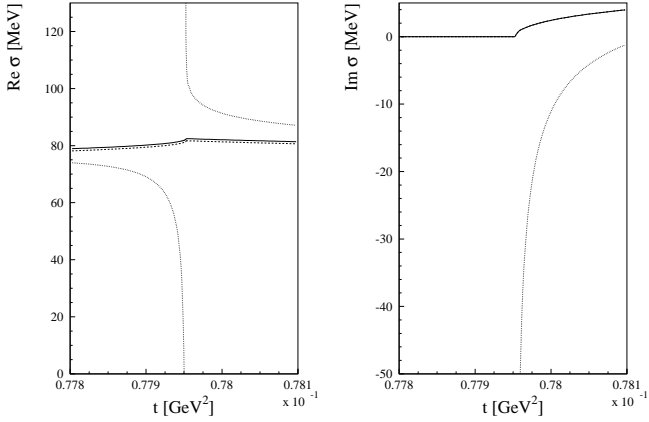}
}
\end{center}
\caption{Real and imaginary parts of the scalar form factor as a function of 
$t$ at ${\cal O}(q^3)$ in the vicinity of $t=4 M_\pi^2$.
Solid lines: EOMS scheme;
dashed lines: infrared regularization (IR) of Ref.\ \cite{Becher:1999he};
dotted lines: HBChPT calculation of Ref.\ \cite{Bernard:1992qa}.
On this scale the (unphysical) divergence of both real and imaginary parts
of the heavy-baryon result becomes visible.
}
\label{fig:scffq3thrnew}       
\end{figure}

   Next we discuss the scalar form factor which is defined as
\begin{displaymath}
\langle N(p')|\hat{m}[\bar{u}(0)u(0)+\bar{d}(0)d(0)]|N(p)\rangle 
=\bar{u}(p')u(p)\sigma(t). 
\end{displaymath}
    The numerical results for the real and imaginary parts of the
scalar form factor at ${\cal O}(q^4)$ are shown in Fig.\ 
\ref{fig:scffq4new} for the extended on-mass-shell scheme
(solid lines) and the infrared regularization scheme (dashed lines).
   While the imaginary parts are identical in both schemes, the
differences in the real parts are practically indistinguishable.
   Note that for both calculations $\sigma(0)$ and $\Delta_\sigma$ have
been fitted to the dispersion results of Ref.\ \cite{Gasser:1990ce}.
   Figure \ref{fig:scffq3thrnew} contains an enlargement near 
$t\approx 4 M_\pi^2$ for the results at ${\cal O}(p^3)$ which
clearly displays how the heavy-baryon calculation 
fails to produce the correct analytical behavior.
   Both real and imaginary parts diverge as $t\to 4 M_\pi^2$.

   As the final example, we consider the electromagnetic form factors of the 
nucleon which are defined via the matrix element of the electromagnetic
current operator as 
\begin{eqnarray*} 
\label{emffdef} 
\lefteqn{\langle N(p_f)\left| J^\mu(0) \right| N(p_i) \rangle=}\nonumber\\ 
&&     \bar{u}(p_f)\left[\gamma^\mu F_1^N(Q^2)+ 
\frac{i\sigma^{\mu\nu}q_\nu}{2m_N}F_2^N(Q^2) \right] u(p_i), \,\, N=p,n, 
\end{eqnarray*} 
where $q=p_f-p_i$ is the momentum transfer and 
$Q^2\equiv-q^2=-t \ge 0$.
   Instead of the  Dirac and Pauli form factors $F_1$ and $F_2$ 
one commonly uses the electric and magnetic Sachs form factors  
$G_E$ and $G_M$ defined by 
\begin{eqnarray*} 
G^N_E(Q^2) &=& F^N_1(Q^2) - \frac{Q^2}{4m_N^2}F_2^N(Q^2),\label{GEN}\\ 
G^N_M(Q^2) &=& F^N_1(Q^2) + F_2^N(Q^2).\label{GMN} 
\end{eqnarray*} 
   At $Q^2=0$, these form factors are given by the electric charges and  
the magnetic moments in units of the charge and the nuclear  
magneton, respectively: 
\begin{eqnarray*} 
&&  G_E^p(0) = 1, \quad G_E^n(0) = 0,\quad
G_M^p(0) = 1+\kappa_p=2.793,\\&& G_E^n(0) = \kappa_n=-1.913. 
\end{eqnarray*} 

\begin{figure}
\begin{center}
\resizebox{0.5\textwidth}{!}{%
\includegraphics{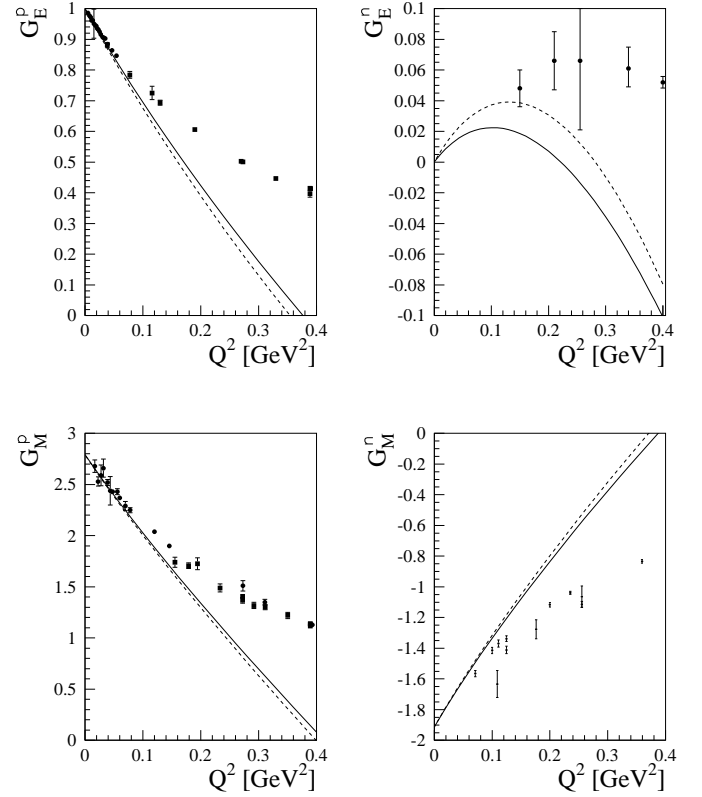}
}
\end{center}
\caption{The Sachs form factors of the nucleon at ${\cal O}(q^4)$. 
The solid and dashed lines refer to the results in the EOMS scheme 
\cite{Fuchs:2003ir} and
the infrared regularization \cite{Kubis:2000zd}, respectively.
   The experimental data for $G_E^p$, $G_E^n$, $G_M^p$, and
$G_M^n$ are taken from Refs.\
\cite{GEP},
\cite{GEN},
\cite{GMP},
and
\cite{GMN}, 
respectively.
}
\label{fig:emffnsachspictpaper}
\end{figure}

   Figure \ref{fig:emffnsachspictpaper} shows the results for the Sachs form 
factors at ${\cal O}(q^4)$ in the EOMS scheme (solid lines) \cite{Fuchs:2003ir}
 and the infrared regularization (dashed lines) \cite{Kubis:2000zd}.
   The description of $G_E^p$, $G_M^p$, and $G_M^n$ turns out to be only 
marginally better than that of the ${\cal O}(q^3)$ calculation 
\cite{Fuchs:2003ir}.
   For the very-small $Q^2$ region the improvement is due to additional 
free parameters which have been adjusted to the magnetic radii.
   As can be seen from Fig.\ \ref{fig:emffnsachspictpaper}, the 
${\cal O}(q^4)$ results only provide a decent description up to 
$Q^2=0.1\,\mbox{GeV}^2$ and do not generate sufficient curvature
for larger values of $Q^2$.
   Moreover, the situation for $G_E^n$ seems to be even worse, where we found
better agreement with the experimental data for the ${\cal O}(q^3)$ results
\cite{Fuchs:2003ir}.
   We conclude that the perturbation series converges, at best, slowly and
that higher-order contributions must play an important role. 

\section{Summary}
\label{summary}
   We have discussed renormalization in the framework of mesonic and
baryonic chiral perturbation theory.
   While the combination of dimensional regularization and the modified
minimal subtraction scheme (of ChPT) leads to a straightforward power 
counting in terms of momenta and quark masses at a 
fixed ratio in the mesonic sector, the situation in the baryonic sector 
proves to be more complicated.
   At first sight, the correspondence between the loop expansion and
the chiral expansion seems to be lost.
   Solutions to this problem have been given in terms of the heavy-baryon
formulation and, more recently, the
infrared regularization approach.

   Here, we have discussed the so-called extended on-mass-shell 
renormalization scheme which allows for a simple and consistent power counting 
in the single-nucleon sector of manifestly Lorentz-invariant chiral 
perturbation theory.
   In this scheme a given diagram is assigned a chiral order $D$ according
to Eq.\ (\ref{dimension1}).
   After reducing the diagram to the sum of dimensionally regularized scalar 
integrals multiplied by corresponding Dirac structures, one
identifies, by expanding the integrands as well as the coefficients in small 
quantities, those terms which need to be subtracted in order
to produce the renormalized diagram with the chiral order $D$ determined 
beforehand.
   Such subtractions can be realized in terms of local counterterms in the 
most general effective Lagrangian.
   Our approach may also be used in an iterative procedure to renormalize 
higher-order loop diagrams in agreement with the constraints due to chiral 
symmetry.
   Moreover, the EOMS renormalization scheme allows for implementing a 
consistent power counting in baryon 
chiral perturbation theory when vector (and axial-vector) mesons
are explicitly included.

%

\end{document}